\documentclass[aps,pra,superscriptaddress,twocolumn]{revtex4-2}
\usepackage{bm}
\usepackage{amsmath}
\usepackage{amssymb}
\usepackage{slashed}
\usepackage{float}
\allowdisplaybreaks
\usepackage{dcolumn}
\newcolumntype{.}{D{x}{}{-1}}
\newcolumntype{d}[1]{D{.}{.}{#1}}
\newcolumntype{w}[1]{D{.}{.}{#1}}
\newcommand*{\centt}[1]{\multicolumn{1}{c}{#1}}

\usepackage{nicefrac}
\usepackage{physics}
\usepackage{graphicx}
\usepackage{xcolor}

\begin{document}
	\title{QED nuclear recoil effect in helium isotope shift}
	
\author{Krzysztof Pachucki}
\affiliation{Faculty of Physics, University of Warsaw,
             Pasteura 5, 02-093 Warsaw, Poland}

\author{Vojt\v{e}ch Patk\'o\v{s}}
\affiliation{Faculty of Mathematics and Physics, Charles University,  Ke Karlovu 3, 121 16 Prague
2, Czech Republic}

\author{Vladimir A. Yerokhin}
\affiliation{Max–Planck–Institut f\"ur Kernphysik, Saupfercheckweg 1, 69117 Heidelberg, Germany}

\begin{abstract}
We present a detailed investigation of the leading-order $m\alpha^5$ QED correction with inclusion
of the finite-nuclear-mass effects. Previously, this correction had been calculated within an expansion
in the electron-nucleus mass ratio $m/M$ up to the first order.
In this work, we derive formulas for the $m\alpha^5$ QED contribution that are valid
up to the second order in $m/M$, and perform its calculation for the $^3$He$-$$^4$He isotope shift,
leading to an improved determination of the nuclear charge-radius difference.

\end{abstract}
\maketitle

\section{Introduction}

The comparison of nuclear charge radii obtained from muonic and electronic atoms provides valuable low-energy tests of precision atomic spectroscopy and of the underlying fundamental interaction theory. 
The ongoing and planned measurements in muonic atoms \cite{nuc_radii} and advances in high-precision laser spectroscopy 
of electronic atoms offer complementary pathways to test QED at unprecedented levels. Combined with increasingly accurate nuclear-structure calculations, the synergy between muonic and electronic systems is expected to deepen our understanding of nuclear structure, ultimately providing more stringent probes of potential physics beyond the Standard Model.

Any persistent discrepancies between nuclear charge radii derived from muonic and electronic-atom spectroscopy
may hint at missing physics or deficiencies in existing theoretical frameworks.
Several such discrepancies have been widely discussed in the past years, but none of them has proven to
be unsolvable within the standard model of fundamental interactions. In particular,
the long-standing proton radius conundrum \cite{pohl:10}  has now been resolved in favor of the $\mu$H value \cite{mohr:22:codata},
not through the discovery of new interactions, but rather through improved measurements in electronic hydrogen \cite{hydrogen:1, hydrogen:2,hydrogen:3}.
A similar discrepancy was reported for the charge-radius difference between the helion and alpha particles, as determined
from muonic and electronic helium spectroscopy \cite{werf:23,schuhmann:23}. However, this problem was also
resolved recently, by identifying a previously overlooked hyperfine-mixing correction
in the theory of electronic helium \cite{qi:24, hfs:sec}. 

In our previous studies \cite{pachucki:15:jpcrd,hfs:sec} we performed a comprehensive analysis of
the $^3$He-$^4$He isotope shift, establishing the theoretical framework for the determination of the nuclear-charge radius difference.
Motivated by the expected experimental progress \cite{eikema:priv}, we now extend our
previous work by calculating the second-order QED nuclear recoil correction 
and thus removing the second-largest uncertainty in the theoretical isotope
shift in helium.

\section{Leading QED in two-body systems}
Before passing to helium, we address first the leading QED contribution of order $m\alpha^5$ for two-body systems consisting 
of a lepton and a nucleus, i.e., hydrogen-like electronic and muonic atoms. We will consider 
the centroid energies, thus neglecting the spin-orbit and tensor spin-spin interactions
which contribute only to the fine and hyperfine structure.
The $m\alpha^5$ QED correction to the energy of a state with
angular orbital momentum $l > 0$ has a simple form \cite{eides}
\begin{align}
E^{(5)} =&\ \label{01}
-\frac{7\,(Z\,\alpha)^5}{6\,\pi}\,\frac{\mu^3}{m_1\,m_2}\,\left\langle \frac{1}{(\mu\,Z\,\alpha\,r)^3}\right\rangle
\\ \nonumber  &\hspace*{-7ex}	
-\frac{2\,\alpha}{3\,\pi}\,\biggl(\frac{1}{m_1}+\frac{Z}{m_2}\biggr)^2
\left\langle\,\vec p\,(H-E)\,\ln\bigg[\frac{2\,(H-E)}{\mu(Z\,\alpha)^2}\bigg]\,\vec{p}\right\rangle
\,,
\\ \nonumber
\end{align}
where the indices 1 and 2 refer to the lepton and the nucleus, respectively, $\mu = m_1\,m_2/(m_1+m_2)$,
$Z$ is the nuclear charge number, $r = |\vec{r}| = |\vec{r}_1-\vec{r}_2|$,  $\vec p = -i\vec{\nabla}$,
the nonrelativistic Hamiltonian $H$ is
\begin{align}
H =&\ \frac{\vec p^{\,2}}{2\,\mu} -\frac{Z\,\alpha}{r}\,,
\label{03}
\end{align}
and $E$ is the reference-state eigenvalue of $H$.
Eq.~(\ref{01}) is valid for arbitrary masses $m_1$ and $m_2$, and the only
approximation involved is the neglect of the nuclear polarizability,
which is considered separately.

We note that Eq.~(\ref{01}) accounts for both the electron and the nucleus self-energy,
the latter given by the term proportional to $(Z/m_2)^2$ in the second line. 
The inclusion of the nuclear self-energy, which is relatively straightforward for the $l>0$ states, 
becomes problematic for the $l = 0$ states, because it also contributes to the nuclear charge radius 
and the nuclear magnetic moment. For this reason, we consider the case of the $l = 0$ states
separately and in more detail. 

Namely, for the $l=0$ states, $E^{(5)}$ acquires extra contact interactions. Assuming a
point-like spin-$1/2$ nucleus, one obtains \cite{eides}
\begin{widetext}
\begin{align}
E^{(5)}(\mathrm{pnt}) =&\
-\frac{7\,(Z\,\alpha)^5}{6\,\pi}\,\frac{\mu^3}{m_1\,m_2}\,\left\langle \frac{1}{(\mu\,Z\,\alpha\,r)^3}\right\rangle		
-\frac{2\,\alpha}{3\,\pi}\,\biggl(\frac{1}{m_1}+\frac{Z}{m_2}\biggr)^2
\left\langle\,\vec p\,(H-E)\,\ln\bigg[\frac{2\,(H-E)}{\mu(Z\,\alpha)^2}\bigg]\,\vec{p}\right\rangle
\nonumber \\ &\
+\frac{(Z\,\alpha)^2}{m_1\,m_2}\,\biggl\{\frac{2}{3}\,\ln(Z\,\alpha)^{-1}+\frac{62}{9}-\frac{2}{m_1^2-m_2^2}\,
\Bigl[m_1^2\,\ln\Bigl(\frac{m_2}{\mu}\Bigr)-m_2^2\,\ln\Bigl(\frac{m_1}{\mu}\Bigr)\Bigr]\biggr\}\,\langle\delta^3(r)\rangle	
\nonumber \\ &\
+\frac{\alpha(Z\alpha)}{m_1^2}\,\bigg(\frac43\ln\frac{m_1\,(Z\alpha)^{-2}}{\mu} + \frac{10}{9} - \frac{4}{15}\bigg)\,\langle\delta^3(r)\rangle
+\frac{Z^2\,\alpha(Z\alpha)}{m_2^2}\,\bigg(\frac43\ln\frac{m_2\,(Z\alpha)^{-2}}{\mu} + \frac{10}{9} - \frac{4}{15}\bigg)\,\langle\delta^3(r)\rangle
\nonumber \\ &\
-\frac{8\,(Z\alpha)^2}{m_2^2-m_1^2}\,\ln\frac{m_2}{m_1}\,\langle\vec s_1\cdot\vec s_2\rangle\,\langle\delta^3(r)\rangle
+\frac{8\,Z\,\alpha^2}{3}\,\frac{\langle\vec s_1\cdot\vec s_2\rangle}{m_1\,m_2}\,\langle\delta^{(3)}(r)\rangle\,,
\label{04}
\end{align}
where $\vec{s}_1$ and $\vec{s}_2$ are the spin operators of the lepton and the nucleus, respectively.
In the above expression,
terms proportional to $(Z\,\alpha)^n$ originate from the two-photon exchange,
those proportional to $\alpha\,(Z\,\alpha)^n$ come from the electron self-energy and vacuum polarization, and those
proportional to $Z^2\,\alpha\,(Z\,\alpha)^n$ are induced by the (point-size) nucleus self-energy and vacuum polarization.
The expectation value of $r^{-3}$ for $l=0$ states is understood as follows
\begin{align}
(\mu\,Z\,\alpha)^3\,\left\langle \frac{1}{(\mu\,Z\,\alpha\,r)^3}\right\rangle
=&\ 4\,\pi\,\lim_{\epsilon\rightarrow 0}\biggl[
 \int dr\, \frac{\phi^2(r)}{r}\,\theta(\mu\,Z\,\alpha\,r-\epsilon) + \phi^2(0)\,\ln(\epsilon)\biggr] \,,
 \label{05}
\end{align}
where $\phi(r)$ is the reference-state wave function.

Let us now rewrite $E^{(5)}$ to the form that could be generalized to an $n$-body system.
The Bethe logarithm can be rewritten as
\begin{align}
\biggl(\frac{1}{m_1}+\frac{Z}{m_2}\biggr)^2
\left\langle\,\vec p\,(H-E)\,\ln\bigg[\frac{2\,(H-E)}{\mu(Z\,\alpha)^2}\bigg]\,\vec{p}\right\rangle
=&\ 	
\left\langle\,\biggl(\frac{\vec p_1}{m_1} -Z\,\frac{\vec p_2}{m_2}\biggr)\,(H-E)\,\ln\bigg[\frac{2\,(H-E)}{\mu\,(Z\,\alpha)^2}\bigg]\,
\biggl(\frac{\vec p_1}{m_1} -Z\,\frac{\vec p_2}{m_2}\biggr)\right\rangle
\,.
\label{06}
\end{align}
Furthermore, we note that although $E^{(5)}$ given by Eq.~(\ref{04}) contains the reduced mass $\mu$, it is in fact independent of $\mu$.
Specifically, the parameter $\mu$ can be replaced by any other mass scale while keeping $m_1$ and $m_2$ unchanged.
This can be demonstrated by the following identity
\begin{align}
(\mu\,Z\,\alpha)^3\,\left\langle \frac{1}{(\mu\,Z\,\alpha\,r)^3}\right\rangle -4\,\pi\,\left\langle \delta^3(r)\right\rangle\,\ln \mu
 =&\
(\mu'\,Z\,\alpha)^3\,\left\langle \frac{1}{(\mu'\,Z\,\alpha\,r)^3}\right\rangle -4\,\pi\,\left\langle \delta^3(r)\right\rangle\,\ln \mu'\,.
\end{align}
and by cancellation of $\ln\mu$ among all terms in Eq. (\ref{04}).
For our purpose, it will be convenient to set the mass scale to $\mu' = m_1$.
Similarly, the $Z$-dependence under the logarithms also cancels out.
Therefore, we rewrite Eq.~(\ref{04}) as
\begin{align}
E^{(5)}(\mathrm{pnt}) =&\
-\frac{14\,(Z\,\alpha)^2}{3\,m_1\,m_2}\, \frac{(m_1\,\alpha)^3}{4\,\pi} \,\left\langle \frac{1}{(m_1\,\alpha\,r)^3}\right\rangle 	
-\frac{2\,\alpha}{3\,\pi}\,\left\langle\,\biggl(\frac{\vec p_1}{m_1} -Z\,\frac{\vec p_2}{m_2}\biggr)\,(H-E)\,\ln\bigg[\frac{2\,(H-E)}{m_1\,\alpha^2}\bigg]\,
\biggl(\frac{\vec p_1}{m_1} -Z\,\frac{\vec p_2}{m_2}\biggr)\right\rangle
\nonumber \\ &\
+\frac{(Z\,\alpha)^2}{m_1\,m_2}\,\biggl[
-\frac{2}{3}\, \ln \alpha +\frac{62}{9}+\frac{2\,m_1^2}{m_2^2-m_1^2}\,\ln\Bigl(\frac{m_2}{m_1}\Bigr)
\biggr]\,\langle\delta^3(r)\rangle	
\nonumber \\ &\
+\frac{\alpha(Z\alpha)}{m_1^2}\,\bigg(\frac43\ln\frac{1}{\alpha^2} + \frac{10}{9} - \frac{4}{15}\bigg)\,\langle\delta^3(r)\rangle
+\frac{Z^2\,\alpha(Z\alpha)}{m_2^2}\,\bigg(\frac43\ln\frac{m_2}{m_1\,\alpha^2} + \frac{10}{9} - \frac{4}{15}\bigg)\,\langle\delta^3(r)\rangle
\nonumber \\ &\
-\frac{8\,(Z\alpha)^2}{m_2^2-m_1^2}\,\ln\frac{m_2}{m_1}\,\langle \vec s_1\cdot\vec s_2 \rangle \,\langle\delta^3(r)\rangle
+\frac{8\,Z\,\alpha^2}{3}\,\frac{\langle \vec s_1\cdot\vec s_2 \rangle }{m_1\,m_2}\,\langle\delta^{(3)}(r)\rangle\,. \label{07}
\end{align}

We now extend our consideration to the case of an arbitrary-spin nucleus with finite size, and drop
all terms  $\propto \vec s_1\cdot\vec s_2$, which contribute to the hyperfine splitting but not to the centroid energy.
The part of the above formula induced by the two-photon exchange $\sim (Z\,\alpha)^2$ was derived for 
the spin-$1/2$ nucleus; it takes a different form for the spin-0 and spin-1 nuclei 
\cite{pachucki:24:rmp}, but this difference is only of order $O(m_1^3/m_2^3)$.
For this reason, we neglect $O(m_1^3/m_2^3)$ terms in the two-photon exchange contribution.
Another problematic set of effects includes the nuclear self-energy (induced by the self-energy loop on the nucleus line)
and the nuclear vacuum polarization, 
since they also contribute  to the nuclear charge radius and magnetic moment. 
These effects have been examined in the literature \cite{radrec}, and a consistent treatment for light 
electronic and muonic atoms has been formulated \cite{pachucki:24:rmp}.
Following this approach, we retain only the logarithmic part of the nuclear self-energy.
Its nonlogarithmic part is absorbed into
the finite nuclear size corrections, discussed in Sec.~\ref{sec:fns}. 
The nuclear vacuum polarization, on the other hand, 
is included into the total hadronic vacuum polarization, which cancels out in the isotope shift.

We thus obtain for centroid energy of hydrogenic systems with an arbitrary-spin nucleus
\begin{align}
E^{(5)} =&\
-\frac{14\,(Z\,\alpha)^2}{3\,m_1\,m_2}\, \frac{(m_1\,\alpha)^3}{4\,\pi} \,\left\langle \frac{1}{(m_1\,\alpha\,r)^3}\right\rangle 	
-\frac{2\,\alpha}{3\,\pi}\,\left\langle\,\biggl(\frac{\vec p_1}{m_1} -Z\,\frac{\vec p_2}{m_2}\biggr)\,(H-E)\,\ln\bigg[\frac{2\,(H-E)}{m_1\,\alpha^2}\bigg]\,
\biggl(\frac{\vec p_1}{m_1} -Z\,\frac{\vec p_2}{m_2}\biggr)\right\rangle
\nonumber \\ &\
+\frac{(Z\,\alpha)^2}{m_1\,m_2}\,\biggl( \frac{1}{3}\, \ln\frac{1}{\alpha^2} +\frac{62}{9} \biggr)\,\langle\delta^3(r)\rangle	
+\frac{\alpha(Z\alpha)}{m_1^2}\,\bigg(\frac43\ln\frac{1}{\alpha^2} + \frac{10}{9} - \frac{4}{15}\bigg)\,\langle\delta^3(r)\rangle
+\frac{Z^2\,\alpha(Z\alpha)}{m_2^2}\,\bigg(\frac43\ln\frac{m_2}{m_1\,\alpha^2}\bigg)\,\langle\delta^3(r)\rangle.
\label{08}
\end{align}

\section{Leading QED in helium atom}
\begin{table*}
\caption{Expansion of the $m\alpha^5$ QED correction in the mass ratio for low-lying states of helium, in kHz.}
\label{tab:one}
\begin{center}
\begin{tabular}{c c | w{10.6}w{6.6}w{6.6}w{10.8}}
\hline
\hline\\[-7pt]
Isotope  & State & \centt{$(m/M)^0$}  & \centt{$(m/M)^1$} & \centt{$(m/M)^2$}   &  \centt{$\Sigma$}\\
\hline\\[-7pt]
${}^3\mathrm{He}$ & $1^1S$ & 40\,506\,157.888 & -13\,730.356 & 17.505 & 40\,492\,445.037\\
			      & $2^1S$ &   2755\,760.767   & -831.835 & 1.153 & 2754\,930.085  \\
				  & $2^3S$ &  3999\,431.448   & -1\,061.422 & 1.394 & 3998\,371.420  \\
				  & $2^1P$ &  38\,769.061     &  624.288 &-0.401 & 39\,392.949 \\
				  & $2^3P$ & -1234\,731.550  & -815.082 & -0.139& -1235\,546.771\\
${}^4\mathrm{He}$ & $1^1S$ &  40\,506\,157.888 & -10\,345.128 & 10.093 & 40\,495\,822.854 \\
				  & $2^1S$ &  2755\,760.767   & -626.746 & 0.665 & 2755\,134.687 \\
				  & $2^3S$ &  3999\,431.448   & -799.728 & 0.805 & 3998\,632.526 \\
				  & $2^1P$ &  38\,769.061     &  470.369 & -0.227& 39\,239.204\\
				  & $2^3P$ &  -1234\,731.550  & -614.123 & -0.083& -1235\,345.756 \\
       \hline \hline
\end{tabular}
\end{center}
\end{table*}

We now turn to generalizing the formulas for the $m\alpha^5$ QED correction obtained in the previous section 
to the case of the helium atom; further extending them to other light atomic systems is straightforward.
In the nonrecoil limit, the expression for the $m\alpha^5$ QED correction is well known \cite{drake:05:springer}.
The first-order recoil correction in $m/M$ was worked out in Ref.~\cite{hel_rec}. Here, we
obtain formulas the $m\alpha^5$ QED correction that include the nuclear recoil effects up to
the second order in the electron-nucleus mass ratio, $(m/M)^2$. As before, we omit terms
of order $(m/M)^3$ and higher, as well as contributions depending on nuclear spin. The finite
nuclear size effects will be addressed in the next sections; for now, we assume the nucleus
to be point-like.

For this generalization of the $m\alpha^5$ QED correction, we use Eq.~(\ref{07}) for the electron-electron terms and Eq.~(\ref{08}) for the electron-nucleus terms,
and assume that there are no three-body terms beyond the Bethe logarithm. The result is
\begin{align}\label{E5he}
E^{(5)} =&\ \delta_1 E^{(5)} + \delta_2 E^{(5)} + \delta_3 E^{(5)} + \delta_4 E^{(5)}\,,
\end{align}
where
\begin{align}
\delta_1 E^{(5)} =&\
-\frac{2\,\alpha}{3\,\pi}\,\biggl(1+Z\,\frac{m}{M}\biggr)^2\,\left\langle\,\frac{\vec p_1+\vec p_2}{m}
\,(H-E)\,\ln\bigg[\frac{2\,(H-E)}{m\,\alpha^2}\bigg]\,\frac{\vec p_1 + \vec p_2}{m}\right\rangle
\nonumber \\ \equiv&\
-\frac{2\,\alpha}{3\,\pi}\,\biggl(1+Z\,\frac{m}{M}\biggr)^2\,
\frac{2\,\pi\,Z\,\alpha}{m^2}\,\left\langle \delta^3(r_1) + \delta^3(r_2)\right\rangle\,\beta\,, \label{10}
\\
\delta_2 E^{(5)} =&\
-\frac{7\,m\,\alpha^5}{6\,\pi} \,\left\langle \frac{1}{(m\,\alpha\,r_{12})^3}\right\rangle
-\frac{7\,m\,Z^2\,\alpha^5}{6\,\pi}\,\frac{m}{M} \,\left\langle \frac{1}{(m\,\alpha\,r_1)^3} + \frac{1}{(m\,\alpha\,r_2)^3}\right\rangle \,, \label{11}
\\
\delta_3 E^{(5)} =&\
\frac{\alpha^2}{m^2}\,\biggl( \frac{14}{3}\, \ln \alpha +\frac{164}{15} \biggr)\,\langle\delta^{(3)}(r_{12})\rangle\,, \label{12}
\\
\delta_4 E^{(5)} =&\
\frac{\alpha^2}{m^2}\,\biggl[
Z\,\frac{4}{3}\,\biggl(\ln\frac{1}{\alpha^2} + \frac{19}{30}\biggr)
+\frac{m}{M}\,Z^2\,\biggl( \frac{1}{3}\, \ln \frac{1}{\alpha^2} +\frac{62}{9}\biggr) + Z^3\,\frac{m^2}{M^2}\,\frac43\ln\frac{M}{m\,\alpha^2}\,
\biggr]\,\langle\delta^3(r_1) + \delta^3(r_2)\rangle\,,
\label{13}
\end{align}
where 
$m$ is the electron mass, $M$ is the nuclear mass, 
the indices 1 and 2 numerate the two electrons, $r_{12} = |\vec r_1 - \vec r_2|$,
the three-particle nonrelativistic Hamiltonian for helium is
\begin{align} \label{eq:15}
H =&\ \frac{\vec p_1^{\,2}}{2\,m} + \frac{\vec p_2^{\,2}}{2\,m} + \frac{(\vec p_1+\vec p_2)^2}{2\,M} + \frac{\alpha}{r_{12}}-\frac{Z\,\alpha}{r_1} -\frac{Z\,\alpha}{r_2}\,,
\end{align}
and the definition of the Bethe logarithm $\beta$ in Eq.~(\ref{E5he}) agrees with that by V. Korobov  in  Ref.~\cite{Korobov:19}.
The expectation values in Eqs.~(\ref{10})-(\ref{13}) are assumed to be evaluated 
with the eigenstates of the three-particle Hamiltonian (\ref{eq:15});
thus, they include the finite nuclear mass effects. 

We have performed numerical calculations of the recoil corrections to all operators in Eqs.~(\ref{10})–(\ref{13}), except for the Bethe logarithm. High-precision numerical values for the Bethe logarithm, including the corresponding recoil corrections of order $m/M$ and $(m/M)^2$, were taken from the work of V. Korobov \cite{Korobov:19}. Our computations of expectation values of various operators were carried out perturbatively in $m/M$, following the numerical approach described in our previous studies \cite{pachucki:15:jpcrd, yerokhin:21:hereview}. Specifically, the expectation value of an arbitrary operator $Q$  was expanded in $m/M$, and terms up to order $(m/M)^2$ were retained,
\begin{align}
\langle Q\rangle  = &\  \langle Q\rangle_{0}
+ \frac{m}{M} \, 2\,\langle Q\,\frac{1}{(E_0-H_0)'}\,\delta_M H\rangle_{0}
+ \left( \frac{m}{M} \right)^2
2\,\Big< Q\,\frac{1}{(E_0-H_0)'}\,(\delta_M H-\langle\delta_M H\rangle_0)\,\frac{1}{(E_0-H_0)'}\,\delta_M H\Big>_{\!0}
 \nonumber \\ &
+ \left( \frac{m}{M} \right)^2
\Big<\delta_M H\,\frac{1}{(E_0-H_0)'}(Q-\langle Q\rangle_{0})\,\frac{1}{(E_0-H_0)'}\,\delta_M H\Big>_{\!0}
\,,
\end{align}
where the subscript ``0'' in $H_0$, $E_0$, and $\langle\ldots\rangle_0$ denotes the infinite-nuclear mass limit,
and $\delta_M H = {\vec P}^2/2 \equiv ({\vec p}_1+{\vec p}_2)^2/2$.

\end{widetext}

Our numerical results obtained for the nonrecoil, leading-order recoil, and second-order recoil corrections
of order $m\alpha^5$ are summarized in Table \ref{tab:one} for the low-lying states of $^3$He and $^4$He.
The nonrecoil and first-order recoil results agree with our earlier work
\cite{pachucki:17:heSummary}, while the second-order recoil results are obtained here for the first time.

Table~\ref{tab:two} presents the individual QED contributions to the ${}^3\textrm{He}$--${}^4\textrm{He}$ isotope shift of
the $2^1S$--$2^3S$ centroid energies.
Most contributions are taken from our previous work \cite{hfs:sec}.
The new result obtained in this study is the $m\alpha^5(m/M)^2$ correction,
which contributes  $-0.101$~kHz to the isotope shift of the $2^1S$--$2^3S$ transition.
This value is twice as large as our earlier estimate of $\pm47$~Hz in 
Ref.~\cite{hfs:sec}.
We note that all recoil effects of order
$m\alpha^2$ and $m\alpha^4$, as well as the $m/M$ recoil correction of order 
$m\alpha^5$ listed in Table~\ref{tab:two}, were recently confirmed by 
independent recalculation in Ref.~\cite{qi:24}.

The dominant uncertainty in the pure QED correction now arises from the unevaluated QED effects of order $m\alpha^7$, 
estimated to be $\pm 0.105$~kHz.
A complete calculation of these contributions is challenging and unlikely to be accomplished in the near future.

\begin{table}
\caption{Pure QED contributions to the ${}^3\textrm{He}$--${}^4\textrm{He}$ isotope shift of
the $2^1S$--$2^3S$ centroid transition frequencies, for the point nucleus, in kHz.
Physical constants are from Ref.~\cite{mohr:22:codata}.}
\label{tab:two}
\begin{center}
\begin{tabular}{c | w{9.5}w{4.3}w{1.3}w{8.6}}
\hline
\hline\\[-7pt]
  & \centt{$(m/M)^1$}  & \centt{$(m/M)^2$} & \centt{$(m/M)^3$}   &  \centt{Sum}\\
\hline\\[-7pt]
 $\alpha^2$          &  -8\,026\,758.512 &     -4\,958.331 & 5.070    & -8\,031\,711.773\\
 $\alpha^4$          &       -2\,496.229 &           2.076      &                &      -2\,494.153\\
 $\alpha^5$          &            56.605 & -0.101                 &                &           56.504\\
 $\alpha^6$          &             2.732  &                            &                 & 2.732 \\
 $\alpha^7$          &        -0.210(105) &                        &                 &             -0.210(105)\\
\hline\\[-7pt]
$E_{\rm qed}$      &&&&    -8\,034\,146.901(105)\\
 \hline \hline
\end{tabular}
\end{center}
\end{table}

\section{Hyperfine mixing effects}

\begin{table}
\caption{Hyperfine mixing contributions to the ${}^3\textrm{He}$$-$${}^4\textrm{He}$ isotope shift of
the $2^1S$--$2^3S$ centroid transition frequencies, in kHz.}
\label{tab:three}
\begin{center}
\begin{tabular}{c | w{4.3}w{4.3}w{4.3}}
\hline
\hline\\[-7pt]
                     & \centt{$(m/M)^2$} & \centt{$(m/M)^3$}   &  \centt{Sum}\\
\hline\\[-7pt]
 ${E}^\mathrm{lo}_\mathrm{mix}$                        &  80.765       & -0.075 &        80.690\\
 $\delta {E}^\mathrm{rel}_\mathrm{mix}$  & 0.137        &               &     0.137  \\
 $\delta {E}^\mathrm{exc}_\mathrm{mix}$  & -1.770       &              &       -1.770\\
\hline\\[-7pt]
$ E_\mathrm{mix}$      &&&    79.056\\
 \hline \hline
\end{tabular}
\end{center}
\end{table}

Among other effects, the hyperfine mixing contribution to the $2^1S$--$2^3S$ transition energy
in $^3$He
requires particular attention because it is enhanced by the small energy separation between the
$2^1S$ and $2^3S$ levels, as first noted by Sternheim \cite{sternheim}. 
This hyperfine mixing correction $E_\mathrm{mix}$ is given by
\begin{align}
E_\mathrm{mix} =&\ \langle H_\mathrm{hfs}\frac{1}{(E-H)'}  H_\mathrm{hfs} \rangle\,,
\end{align}
where $H_\mathrm{hfs}$ is the leading-order effective Hamiltonian responsible for the hyperfine
structure, see Ref.~\cite{hfs:sec} for details.

The leading-order contribution is due to the mixing between the $2^3S_1$ and $2^1S_0$ states
and is given by
\begin{align}\label{eq:18}
E_\mathrm{mix}^{\mathrm{lo}} = \frac{\left|\langle 2^3S|H_\mathrm{hfs}|2^1S\rangle_0\right|^2}{E_0(2^1S)-E_0(2^3S)}\,,
\end{align}
where the superscript ``0'' indicates the nonrecoil limit. The leading-order term was taken into
account already in our earlier works \cite{pachucki:15:jpcrd, pachucki:17:heSummary}.

The recoil correction to $E_\mathrm{mix}^{\mathrm{lo}}$
accounts for the finite nuclear mass in the matrix element of $H_\mathrm{hfs}$ and
in the energy denominator. For its calculation we use our
result for the matrix element of the Fermi contact interaction for $^3$He
\begin{equation}
	4\pi\,\langle 2^3S|\delta^3(r_1)-\delta^3(r_2)|2^1S\rangle = 29.118\,9786\,,
\end{equation}
which exactly includes the finite nuclear mass. For comparison, this matrix element
in the infinite nuclear mass limit is
\begin{equation}
	4\pi\,\langle 2^3S|\delta^3(r_1)-\delta^3(r_2)|2^1S\rangle_0 = 29.134\,978\,.
\end{equation}

The relativistic correction to $E_\mathrm{mix}^{\mathrm{lo}}$
comes from the relativistic shift of the $2^3S$--$2^1S$ energy difference,
as well as the electron anomalous magnetic moment (amm) and the nuclear-structure
corrections,
\begin{equation}
\delta E^\mathrm{rel}_\mathrm{mix}	= E_\mathrm{mix}^{\mathrm{lo}}
\Big[(1+\kappa + \delta_\mathrm{nuc})^2 - \frac{\delta E_\mathrm{rel}}{\delta E} - 1 \Big]\,,
\end{equation}
where $\kappa$ is the electron amm, 
$\delta_\mathrm{nuc}$ is the nuclear-structure contribution taken
from Ref.~\cite{patkos:23:heplus}, and 
$\delta E_{\rm rel}$ is the relativistic correction to $\delta E  = E_0(2^1S)-E_0(2^3S)$, see also Ref.~\cite{hfs:prl}.

The next important correction $E_\mathrm{mix}^{\mathrm{exc}}$ is due to the hyperfine mixing with the
$n>2$ excited states.
Its significance was first pointed out in Ref. \cite{qi:24}. In our previous work \cite{hfs:sec} we verified 
it and accurately calculated this correction.
Table~\ref{tab:three} summarizes our numerical results obtained for individual hyperfine-mixing corrections.

\section{Nuclear size effects}
\label{sec:fns}

\begin{table}
\caption{Nuclear polarizability and higher-order nuclear size corrections to 
the $^3$He--$^4$He isotope shift of the $2^1S$--$2^3S$ transition, in kHz.
}
\label{tab:four}
\begin{center}
\begin{tabular}{c | w{3.6}w{3.6}w{3.6}w{3.6}}
\hline
\hline\\[-7pt]
Contribution  & \centt{$(m/M)^0$}  & \centt{$(m/M)^1$} &  \centt{Sum}\\
\hline\\[-7pt]
${  E}_\mathrm{pol}$  &             0.198(20)&  &  0.198(20)\\
${  E}_\mathrm{fns}^{(5)}$ & 0.045 & 0.004 & 0.049 \\
${  E}_\mathrm{fns}^{(6)}$ & -0.461 & 0.003 & -0.458 \\
${  E}_\mathrm{radfns}^{(6)}$ & 0.054 & & 0.054\\ \hline
$\Sigma$ & & & -0.157(20) \\ \hline \hline
\end{tabular}
\end{center}
\end{table}

The leading finite nuclear size (fns) correction to an energy level is of order $m\alpha^4$ and is given by
\begin{align}
E^{(4)}_\mathrm{fns}[{}^A\textrm{He}] =&\
	\frac{2\pi}{3}\, Z\,\alpha^4\,m\,
	\phi^2(0)\,\frac{r^2_C}{\not\!\lambda^2} \equiv C_A\, r^2_C
\,, \label{86}
\end{align}
where $\phi^2(0) = \sum_a \langle\delta^3(r_a)\rangle$,
$r_C$ is the root-mean-square charge radius of the nucleus,
$\not\!\!\!\lambda = 386.159$ fm is the reduced Compton wavelength of the electron, $A$ is
the isotope mass number, and
the expectation value of the $\delta$-function includes finite nuclear mass effects.

As we pointed out in our previous work \cite{hfs:sec},
because of the mass dependence,
the coefficient $C_A$ in the above equation depends (weakly) on the isotope $A$.
For this reason, we write the fns contribution
to the  ${}^3\textrm{He}$--${}^4\textrm{He}$ isotope shift as \cite{hfs:sec}
\begin{align}
E^{(4)}_\mathrm{fns}[{}^3\textrm{He}\!-\!{}^4\textrm{He}] =&\ C_3\,r^2_3 - C_4\,r^2_4
\nonumber \\ =&\
C\,\big[r^2_3 - r^2_4\big] + D\,\big[r^2_3 + r^2_4\big]\,, \label{87}
\end{align}
where $r_A \equiv r_C(^A\textrm{He})$, and the last line is the definition of the coefficients $C$ and $D$.

There are numerous higher-order fns corrections, investigated in detail in 
Refs.~\cite{pachucki:18, pachucki:23, pachucki:25a, pachucki:25b}.
Specifically, the $m\,\alpha^5$ nonrecoil fns correction is given by
\begin{align}
E^{(5,0)}_\mathrm{fns}  =&\ -\frac{\pi}{3}\,\phi^2(0)\,(Z\,\alpha)^2\,m\,r^3_F\,,
\end{align}
where
$r_F$ is the Friar radius, which for the  exponential (dipole) parametrization of the nuclear-charge distribution
is given by $r_F = 1.558\,965\,r_C$.
The recoil $m\alpha^5$ fns correction for the exponential nuclear-charge distribution is given by \cite{pachucki:23}
\begin{align}
E^{(5,1)}_\mathrm{fns}  =& -\frac{\phi^2(0)}{M\,m}\,(Z\,\alpha)^2\biggl(-\frac{43}{12} +\ln 12 -2\,\ln m\,r_C\biggr) m^2 r_C^2.
\end{align}
The next-order in $\alpha$ correction
$E^{(6,0)}_\mathrm{fns}$ is known only for hydrogenic systems and is state dependent \cite{pachucki:18}.
Since a large part of  this correction scales with $\phi^2(0)$, we generalize it to many-electron systems
by using the hydrogenic result for $n=1$,
\begin{align}
E^{(6,0)}_\mathrm{fns} \approx& -(Z\,\alpha)^3\,r_C^2\,\frac{2\,\pi}{3}\,\phi^2(0)\,\Big[\ln(m\,r_C\,Z\,\alpha) - 0.413\,384\Big] .
\end{align}
The recoil fns correction $E^{(6,1)}_\mathrm{fns}$ in the dipole parametrization is given by \cite{pachucki:25a}
\begin{align}
E^{(6,1)}_\mathrm{fns} =&\ -\frac{\pi}{M}\,(Z\,\alpha)^3\,\phi^2(0)\,{r_C}\,0.962\,211\,.
\end{align}
Finally, the radiative fns correction is \cite{eides}
\begin{align}
E^{(6,0)}_\mathrm{radfns} =&\ \alpha\,(Z\,\alpha)^2\,\frac{\phi^2(0)}{m^2}\,\frac{2\,\pi}{3}\,(m\,r_C)^2\,(4\,\ln2-5)\,.
\end{align}
Further fns corrections are of higher orders in the mass ratio and/or the fine structure constant  $\alpha$.
They are negligibly small for helium \cite{pachucki:25b}.

Apart of the nuclear size, one must also to account for the nuclear polarizability correction $E_\mathrm{pol}$.
The leading-order nuclear polarizability of order $m\alpha^5$ comes from the two photon exchange
and was calculated in Refs.~\cite{pachucki:07:heliumnp, muli:24}.

Table~\ref{tab:four} summarizes our numerical
results for the higher-order fns  and nuclear polarizability corrections for the $^3$He--$^4$He isotope 
shift of the $2^1S$--$2^3S$ transition. The fns corrections were calculated with the following values
of the nuclear charge radii: 
$r_C(^3\mathrm{He}) = 1.678\,6(12)\,\mathrm{fm}$ and $r_C(^4\mathrm{He}) = 1.970\,07(94)\,\mathrm{fm}$
\cite{pachucki:24:rmp}.
Numerical values of the coefficients $C$ and $D$ in Eq.~(\ref{87}) are listed
in Table~\ref{tab:five}.

\section{Charge radii difference}

We are now in a position to determine the difference of the mean square charge radii of the helium isotopes,
$\delta r^2 = r^2_C(^3\mbox{\rm He}) - r^2_C(^4\mbox{\rm He})$.
Table~\ref{tab:five} summarizes all experimental and theoretical input required for this determination.
The $2^1S$--$2^3S$ transition energy in $^4$He was measured in Ref.~\cite{rengelink:18}.
To obtain the corresponding centroid energy in $^3$He, we combine the
$2^1S^{F=1/2}$--$2^3S^{F=3/2}$ transition energy measured in Ref.~\cite{werf:23} with
the known experimental hyperfine-structure interval of the $2^3S^{F=3/2}$  state \cite{schluesser:69,rosner:70}.
The experimental centroid-energy isotope shift is combined with the QED theory predictions summarized in 
Tables~\ref{tab:two}-\ref{tab:four}. 
The remainder is attributed to the leading-order
fns contribution given by Eq.~(\ref{86}), from which the charge radii difference $\delta r^2$ is
determined. We note that although
the higher-order fns corrections summarized in
Table~\ref{tab:four} depend on the nuclear charge radii, these corrections are sufficiently small that
the uncertainties of the existing values of the nuclear-charge radii do not contribute at the level of
our interest.

Our result for the mean square charge radius difference, $\delta r^2 =  1.0679\,(13)$ fm$^2$,
agrees within $1.3\,\sigma$ with the value of $1.0636\,(31)$ fm$^2$ derived from the muonic helium
\cite{schuhmann:23}.
It should be mentioned that in our previous work \cite{hfs:sec} there was a mistake in
evaluation of the uncertainty of $\delta r^2$. Consequently, the uncertainty of 
$\pm 0.0007$ fm$^2$ printed in Ref.~\cite{hfs:sec} should be replaced by $\pm 0.0014$ fm$^2$.

\section{Summary}
We have derived a formula for  the second-order recoil correction to the leading QED contribution, and performed a calculation for the helium atom. 
This calculation removed the second-largest theoretical uncertainty in the isotope shift of the $2^1S$--$2^3S$ transition. 
Using the updated QED theory together with the available experimental transition energies, we determined the mean-square charge radius difference 
$\delta r^2$ between the helium isotopes. Our result agrees with the value derived from muonic helium \cite{schuhmann:23} at the $1.3\,\sigma$ level, 
while being 2.4 times more precise. The small deviation from the muonic-helium value may stem from nuclear-polarizability effects, which limit the
theoretical accuracy in the muonic helium Lamb shift. 

An important advantage of determining $\delta r^2$ from electronic helium, as compared 
with muonic helium, is its lower sensitivity to nuclear polarizability effects.
As a consequence, the uncertainty of the electronic $\delta r^2$ value arising from the nuclear
polarizability is just $0.0001$~fm$^2$, whereas in the muonic helium it is 30 times larger.

At present, the limiting factor in the determination of $\delta r^2$ 
from the electronic helium
is the experimental accuracy \cite{werf:23, rengelink:18}. In the future, upcoming experiments aim to improve the precision of the $2^1S$--$2^3S$ transition energy to about 50 Hz \cite{eikema:priv}, which would reduce the total uncertainty in $\delta r^2$ to $0.0005$ fm$^2$.

Once this is accomplished, any further improvement in the accuracy of $\delta r^2$  would
require a complete calculation of the $m\alpha^7$ QED recoil effect.
This would be a significant challenge, as these effects are currently
unknown even for hydrogenic systems. 
Nevertheless, such a calculation is possible at least in principle, 
in contrast to major further advances in the theory of nuclear polarizability, which 
limits the $\delta r^2$ determination in muonic helium.

\begin{table*}
\caption{Determination of the $^3$He\,--\,$^4$He nuclear charge difference $\delta r^2$ from the isotope shift of the
$2^1S$\,--\,$2^3S$ transition,
in kHz unless specified otherwise. Physical constants are from Ref. \cite{mohr:22:codata}.}
\label{tab:five}
\begin{ruledtabular}
  \begin{tabular}{l.l}
$E(^3{\rm He},2^1S^{F=1/2} - 2^3S^{F=3/2})$ &  192\,504\,914\,418x.96(17) &Experiment \cite{werf:23}\\
$-E(^4{\rm He},2^1S - 2^3S)$ & -192\,510\,702\,148x.72(20) & Experiment \cite{rengelink:18} \\
$\delta E_{\rm hfs}(2^3S^{3/2})$& -2\,246\,567x.059(5) & Experiment \cite{schluesser:69,rosner:70}\\
$-\delta E_{\rm iso}(2^1S - 2^3S)$ (QED, point nucleus) &8\,034\,146x.901\,(105) & Theory, Table~\ref{tab:two} \\
$-\delta E_{\rm iso}(2^1S - 2^3S)$ (hyperfine mixing) &-79x.056 & Theory, Table~\ref{tab:three} \\
$-\delta E_{\rm iso}(2^1S - 2^3S)$ (nuclear structure) &0x.157\,(20) & Theory, Table~\ref{tab:four} \\ [1ex]
Sum               &-228x.82\,(26)_{\rm exp}(11)_{\rm the} & \\
$C$                            &-214x.353\,\,\, {\rm kHz/fm}^2  &  \\
$D$                            &0x.013\,\,\, {\rm kHz/fm}^2  &  \\[1ex]
$\delta r^2 = r^2_C(^3\mbox{\rm He}) - r^2_C(^4\mbox{\rm He})$                   & 1x.0679\,(12)_{\rm exp}(5)_{\rm the}\;{\rm fm}^2             & this work\\
& 1x.0636\,(6)_\mathrm{exp}(30)_\mathrm{the}\;{\rm fm}^2 & $\mu^{3,4}$He$^+$ Lamb shift \cite{schuhmann:23} \\
  \end{tabular}
\end{ruledtabular}
\end{table*}

\end{document}